# A BI-LEVEL APPROACH FOR OPTIMAL CONTRACT PRICING OF INDEPENDENT DISPATCHABLE DG UNITS IN DISTRIBUTION NETWORKS

*Ashkan Sadeghi Mobarakeh, Abbas Rajabi-Ghahnavieh\*, Hossein Haghighat\*\**
*\*Sharif University of Technology*
*\*\* University of Jahrom*
*E-mail: AshkanSdeghi85@gmail.com*

**ABSTRACT**

Distributed Generation (DG) units are increasingly installed in the power systems. Distribution Companies (DisCo) can opt to purchase the electricity from DG in an energy purchase contract to supply the customer demand and reduce energy loss. This paper proposes a framework for optimal contract pricing of independent dispatchable DG units considering competition among them. While DG units tend to increase their profit from the energy purchase contract, DisCo minimizes the demand supply cost. Multi-leader follower game theory concept is used to analyze the situation in which competing DG units offer the energy price to DisCo and DisCo determines the DG generation. A bi-level approach is used to formulate the competition in which each DG problem is the upper-level problem and the DisCo problem is considered as the lower-level one. Combining the optimality conditions of all upper-level problems with the lower level problem results in a multi-DG equilibrium problem formulated as an equilibrium problem with equilibrium constraints (EPEC). Using a nonlinear approach, the EPEC problem is reformulated as a single nonlinear optimization model which is simultaneously solved for all independent DG units. The proposed framework was applied to the Modified IEEE 34-Bus Distribution Test System. Performance and robustness of the proposed framework in determining econo-technically fare DG contract price has been demonstrated through a series of analyses.

*Keywords:* Distributed Generation (DG), Distribution Company (DisCo), Pricing, Game Theory, Multi-leader follower, Equilibrium Problems with Equilibrium Constraints (EPEC).

Nomenclature

### A. *Indexes and Sets*

| | |
|---|---|
| $k, l$ | Node number indices. |
| $i$ | Distributed generation unit index. |
| $kl$ | Composed index for a line between nodes $k$ and $l$. |
| $\mathbf{K}$ | Set of nodes. |
| $\mathbf{L_k}$ | Set of nodes connected to node $k$. |
| $\mathbf{I}$ | Set of leaders. |
| $\mathbf{T}$ | Set of time period |
| $S$ | Set of slack variables s |
| $K', K'', K''', K'''', K''''', K''''''$ | Subsets of S |

### B. *Parameters*

| | |
|---|---|
| $P_{dk}(t)$ | Active power demand at node $k$ in period $t$ (MW). |
| $Z_{kl}$ | Impedance magnitude for the line connecting nodes $k$ and $l$ ($\Omega$). |
| $\overline{P}_{kl}$ | Maximum active power limit in the line connecting nodes $k$ and $l$ (MW). |
| $\underline{V}_k$ | Minimum voltage magnitude limit at node $k$ (V). |
| $\overline{V}_k$ | Maximum voltage magnitude limit at node $k$ (V). |
| $\underline{P}_{sb}$ | Minimum active power limit at the substation in period $t$ (MW). |
| $\overline{P}_{sb}$ | Maximum active power limit at the substation in period $t$ (MW). |
| $\underline{P}_{dg_i}$ | Minimum active power limit of DG unit $i$ (MW). |
| $\overline{P}_{dg_i}$ | Maximum active power limit of DG unit $i$ (MW). |
| $c_i$ | Production cost of DG unit $i$ (€/MWh). |
| $\beta(t)$ | Energy price at the substation in period $t$ (€/MWh). |

### C. *Variables*

| | |
|---|---|
| $\alpha_i$ | Contract price of DG unit $i$ (€/MWh). |
| $X_i$ | Decision variable of DG$i$ |
| $P_{dg_i}(t)$ | Active power generated by DG unit $i$ in period $t$ (MW). |
| $P_{sb}(t)$ | Active power purchased at the substation in period $t$ (MW). |
| $P_{gk}(t)$ | Active power injection in node k in period $t$ (MW). |
| $V_k(t)$ | Voltage magnitude at node $k$ in period $t$ (V). |
| $\lambda_k(t)$ | Dual variable related to the power balance constraint at node $k$ in period $t$. |
| $\overline{\mu}_{kl}(t)$ | Dual variable for maximum flow constraint of line connecting nodes $k$ and $l$ in at $t$. |
| $\underline{\mu}_{kl}(t)$ | Dual variable for minimum flow constraint of line connecting nodes $k$ and $l$ in period $t$. |
| $\overline{\mu}_k^v(t)$ | Dual variable related to the constraint on maximum level of $V_k(t)$. |
| $\underline{\mu}_k^v(t)$ | Dual variable related to the constraint on minimum level of $V_k(t)$. |
| $\overline{\mu}_{sb}(t)$ | Dual variable related to the constraint on maximum level of $P_{sb}(t)$. |
| $\underline{\mu}_{sb}(t)$ | Dual variable related to the constraint on minimum level of $P_{sb}(t)$. |
| $\overline{\mu}_i^{dg}(t)$ | Dual variable related to the constraint on maximum level of $P_{dg_i}(t)$. |
| $\underline{\mu}_i^{dg}(t)$ | Dual variable related to the constraint of minimum level of $P_{dg_i}(t)$. |

### D. *Symbols*

| | |
|---|---|
| $a \perp b$ | Complementarity condition between $a$ and $b$. |

### E. Vectors

| | |
|---|---|
| w | Vector consists of variables Pdg, Psb and V. |
| $\Lambda$ | Dual variables of the equality constraints in DisCo optimization problem. |
| $y_1$ | Vector consists of w and $\lambda$. |
| M | Dual variables for the inequality constraints in DisCo optimization problem |
| $h_e$ | All the equality constraints in DG$_i$ optimization problem. |
| $\bar{\mu}_i$ | Complementarity variables for $h_e$ in DG$_i$ optimization problem. |
| $h_{in}$ | All the inequality constraints in DG$_i$ optimization problem. |
| $\underline{\mu}_i$ | Complementarity variables corresponding to $h_{in}$ in DG$_i$ problem. |
| $\psi_i$ | Complementarity variables for $-\mu s \geq 0$ constraints in DG$_i$ optimization problem. |
| $\sigma_i$, | Complementarity variables for $s \geq 0$ constraints in DG$_i$ optimization problem. |
| $\phi_i$ | Complementarity variables for $\mu \geq 0$ constraint in DG$_i$ optimization problem. |

F. Others

| | |
|---|---|
| $f_{DisCo}$ | Objective function of DisCo |
| $f_i$ | Objective function of DG$i$ |

## 1 INTRODUCTION

Over the last decade, there has been a growing trend toward Distributed Generation (DG) application in the distribution networks. DG has received great interests due to its potential benefits to both owner and power system such as energy revenue, energy loss reduction, voltage profile enhancement and deferral of distribution network expansion [1-2]. Restructuring in the electrical industry, technology advancement as well as environmental concerns have also caused DG to be even more attractive.

One important goal in power system restructuring is to provide the opportunity for independent commercial entities to participate in power generation or electricity distribution. So, independent Distribution Company (DisCo) and DG units can be found in a deregulated environment in which each entity tries to optimize its own objective.

To supply the demand of customers, DisCo purchases energy from the electricity market or through bilateral contracts with generation companies. Electric energy is injected to the distribution network through the sub-transmission substations. Although in most cases, DG units cannot compete with power plants, they can participate in providing energy for the distribution networks. Therefore, DisCo might opt to purchase energy from DG units located in the distribution network as well. Since the potential benefit of DG for the distribution network depends largely on the DG location, size and generation, various studies have been reported to determine the optimal size and place of DG units [3-5]. A comprehensive study on optimal DG placement was conducted in [6] and [7]. Reference [8] quantified DG benefits to DisCo and the electricity market and proposed a method to reward DG units according to their benefits.

As there might be conflict of interest between DisCo and DG units, appropriate approach is needed to determine a proper price for energy generated by DGs in the distribution networks. The nodal pricing for DG is formulated as the least-cost dispatch problem in [9]. Reference [10] has used locational marginal price (LMP) approach for DG pricing considering load uncertainties. Some related works in this area are available in [11]. DG generation pricing when DG is owned by DisCo has been studied in [12]. Role of retailer-owned DG unit to improve the profit of the retailer has been discussed in [13]. Most available studies have been developed from DisCo standpoint and have focused to maximize the DG benefit to DisCo.

To the authors' knowledge, few recent studies have simultaneously considered the interests of both DisCo and DG units as independent entities. Game theory provides a conceptual framework to address the problems in which the interest of more than one entity (player) is to be considered. In [14], optimal DG location and contract pricing is determined. A bilevel optimization method has been used in [15] to find the DG optimal contract price for DisCos to dispatch DG resources. In these works, the total profits of DGs belonging to one DG owner are maximized while the DisCo payment is minimized using a mathematical program with equilibrium constraints (MPEC). Using an iteration method, [16] determined the optimal contract pricing for one private DG owner. However, the methods presented in [14-16] are limited to consider the competition between DisCo and a single DG owner, forming a two-player game, and the methods are not applicable to the cases where there are more than two players compete in the distribution network.

The competition can get more sophisticated as two or more than two independent DG units or electricity supplier/retailer participate in DisCo supply basket. On the other side, one of most convenient electricity businesses is investment in DG units. More and more DG units are expected to be installed in the distribution networks making a competitive environment between DG owners. A framework is, therefore, needed to determine optimal DG contract price taking into account the competition among independent DG owners in the distribution network.

In this paper, a framework is proposed for optimal contract pricing of competing dispatchable DG units in the distribution system. A multi-leader follower game theory concept is used in which the DG units are considered as leaders and DisCo is follower. DG objective is to maximize the electricity generation profit while the DisCo objective is to minimize the electricity supply cost. DG units give the price offer to DisCo and DisCo determines the amount of energy which should be purchased from the wholesale electricity market and DG units for the different periods of the contract between DisCo and DG units.

The problem is formulated as a bi-level optimization problem in which each DG optimization problem forms the upper-level problem that should be solved subjected to the lower-level problem. DisCo problem, on the other hand, is considered as the lower-level problem which is represented as an optimal power flow problem. For each DG, in such setting, a mathematical program with equilibrium constraint (MPEC) is obtained. By combining the optimality conditions of problems of all DG units, an equilibrium problem with equilibrium constraints (EPEC) is formed. The EPEC is then reformulated as a nonlinear programming (NLP) model using an NLP method. The proposed framework has been applied to the 3-bus and Modified IEEE 34-Bus Distribution Test System and the performance of the proposed framework has been evaluated.

In their previous work [17], the authors have presented a diagonalisation technique to determine DG optimal contract price considering DG units competition. However, the diagonalization technique is time-consuming and faces cycling even for the small 6-bus test system used in [17] and is not applicable for practical cases. In the present paper, we extended our previous work to be applicable in practical cases. Proposed NLP formulation can solve all DG problems simultaneously without facing cycling. Insightful and interesting case studies are presented in this work. The importance of the competition is also presented in detail. We also compared our work with the pervious related research [15], [16] and the optimality of the results were tested using two techniques.

Main contributions of this paper are:

1) It presents a mathematical NLP model for obtaining optimal contract pricing of non-utility dispatchable DG units in the distribution system.
2) It explicitly models and quantifies the impact of the DG interaction on the solution using a game theoretic approach and a multi-period bilevel optimization model.
3) The reaction of DisCo to the energy contract prices offered by independent DG units is explicitly accounted for.

The paper is arranged as follows: Section 2 describes the structure of the problem, and the problem formulation. Solution method is presented for a simple distribution system in Section 3. Case studies are presented and discussed in Section 4. Optimality evaluation and computational issues are evaluated in Section 5. Practical implications of the proposed framework are discussed in Section 6 and the concluding remarks are provided in Section 7.

## 2 MATHEMATICAL FORMULATION

### 2.1 Basic Assumptions

It is assumed that the distribution system is operated by a single DisCo. The distribution system is connected to the subtransmission substatiuon. Independent investor-owned DG units are assumed to be installed in the distribution system which are based on dispatchable technology. DisCo can provide the electricity from the wholesale market or through bilateral contracts with DG units. A market structure similar to the ones presented in [14-17] is considered in which DisCo is to provide costumers' demand at minimum operation cost and power loss.

DG units compete to increase their profit by increasing the amount or price of the generation contracted with DisCo. On the other hand, DisCo takes benefits from DG generation to improve the voltage profile, reduce power loss and supply the demand. The proposed multi-leader follower game-theory framework to model the competition among DG units and DisCo is presented in Fig. 1.

**Figure 1:** Schematic of the Proposed Framework.

It can be seen in Fig. 1 that, in the proposed framework, first, independent DG units anticipate the DisCo reaction and compete with each other. They playing a Nash game to offer their own contract prices and available capacities to DisCo. Then DisCo determines the quantity of energy which should be purchased from DG units and electricity market by minimizing total DisCo cost, considering the network constraints.

From the regularity point of view, the proposed framework is appropriate for any electricity market in which power loss are assumed by the DisCo. This is the case of most countries with deregulated electricity markets, such as U.K., in which incentive mechanisms imposed by the regulator force Discos to reduce (or assume) power losses. In most power market, such as Iran Power Market [18], DisCo takes both technical and financial responsibilities for distribution system operation and also for buying the electricity directly from the DG units [19].

To determine the amount of energy to be purchased from each supplier, DisCo must weigh not only the energy prices, but also it must consider the impact of the supplied energy on the distribution network including loss reduction and voltage profile improvement. We take into account these two aspects by means of an approximate power flow model as in [20] in which the active power flow between node k and l is approximated as follows:

$$P_{kl} \cong V_k \cdot \frac{(V_k - V_l)}{Z_{kl}} \tag{1}$$

### 2.2 Bilevel Model

The problem of each DG unit can be formulated as a bilevel problem considering two decision makers referred to as leader and follower that will try to optimize their individual objective functions. The DG, in the upper level, offers the contract price to the DisCo over a specified time period so as to maximize its own profit. The DisCo, in the lower level, minimizes total energy supply cost and determines the power purchased from DG.

The problem faced by each DG unit can be formulated as the following bilevel problem:

$$\underset{\alpha_i}{\text{Max}} \quad \sum_{t \in T} (\alpha_i - c_i) P_{dg_i}(t) \tag{2}$$

Subject to:

$$\underset{V, P_{dg_i}, P_{sb}}{\text{Min}} \quad \sum_{t \in T} \beta(t) P_{sb}(t) + \sum_{t \in T} \sum_{i \in I} \alpha_i P_{dg_i}(t) \tag{3}$$

Subject to:

$$-P_{gk}(t)+P_{dk}(t)+\sum_{l \in L_k} \frac{V_k(t)(V_k(t)-V_l(t))}{Z_{kl}} = 0 \quad :\lambda_k(t); \quad \forall k \in K, \forall t \in T; \quad (4)$$

$$-\overline{P}kl \leq \frac{V_k(t)(V_k(t)-V_l(t))}{Z_{kl}} \leq \overline{P}kl \quad :\overline{\mu}_{kl}(t), \underline{\mu}_{kl}(t); \forall k \in K, \forall t \in T; \quad (5)$$

$$\underline{V}_k \leq V_k(t) \leq \overline{V}k \quad :\overline{\mu}_k^v(t), \underline{\mu}_k^v(t); \forall k \in K, \forall t \in T; \quad (6)$$

$$\underline{P}_{sb} \leq P_{sb}(t) \leq \overline{P}_{sb} \quad :\overline{\mu}_{sb}(t), \underline{\mu}_{sb}(t); \quad \forall t \in T; \quad (7)$$

$$\underline{P}_{dg_i} \leq P_{dg_i}(t) \leq \overline{P}_{dg_i} \quad :\overline{\mu}_i^{dg}(t), \underline{\mu}_i^{dg}(t); \quad \forall t \in T, \forall i \in I. \quad (8)$$

In which:

Equation (2) represents DG objective function;
Equation (3) reflects DisCo objective function;
Equation (4) represents bus power balance;
Equation (5) represents distribution line flow limit;
Equation (6) gives bus voltage limit;
Equation (7) represents the limit on the subtransmission substation power;
Equation (8) gives the limit on different DGs' generations.

Note that in equations (4) to (8) and for most of constraint equations in this paper, the dual variable associated with the equation is mentioned after ":" symbol and the complementary variable associated with each equation is mentioned after "⊥" symbol.

The set of equations (3)-(8) form the lower level problem, i.e. DisCo problem. DisCo objective function comprises of two terms: the cost of power purchased from the market at a price of β(t); and the cost of energy purchased from all DG units at their contract prices. Equation (2) forms the upper level problem that represents DG objective function. The upper level problem determines the optimal contract pricing of each DG unit over the contract period whereas the dispatch level of DG and wholesale market purchases result from the solution of the lower level problem. Using the KKT conditions of the lower level, each DG problem can be cast as follow:

$$Max_{\alpha_i, v_k, P_{dg_i}, P_{sb}} \sum_{t \in T} (\alpha_i - c_i) P_{dg_i}(t) \quad (9)$$

Subject to:

$$\nabla_{v_k(t)} \rightarrow (-\lambda_k(t) + \bar{\mu}_{kl}(t) - \underline{\mu}_{kl}(t)) \sum_{l \in L} \frac{(2v_k(t)-v_l(t))}{z_{kl}} + \bar{\mu}_k^v(t) - \underline{\mu}_k^v(t) = 0; \forall k \in K, \forall t \in T; \quad (10)$$

$$\nabla_{P_{dg_i}(t)} \rightarrow \alpha_i + \lambda_k(t) + \bar{\mu}_i^{dg}(t) - \underline{\mu}_i^{dg}(t) = 0; \forall t \in T, \forall i \in I; \quad (11)$$

$$\nabla_{P_{sb}(t)} \rightarrow \beta(t) + \lambda_k(t) + \bar{\mu}_{sb}(t) - \underline{\mu}_{sb}(t) = 0; \forall t \in T; \quad (12)$$

$$-P_{gk}(t) + P_{dk}(t) + \sum_{l \in L} \frac{v_k(t)(v_k(t)-v_l(t))}{z_{kl}} = 0; \forall k \in K, \forall t \in T; \quad (13)$$

$$\sum_{l \in L} \frac{v_k(t)(v_k(t)-v_l(t))}{z_{kl}} + \bar{P}_{kl} \geq 0 \perp \underline{\mu}_{kl}(t) \geq 0; \forall k \in K, \forall t \in T; \quad (14)$$

$$-\sum_{l \in L} \frac{v_k(t)(v_k(t)-v_l(t))}{z_{kl}} + \bar{P}_{kl} \geq 0 \perp \bar{\mu}_{kl}(t) \geq 0; \forall k \in K, \forall t \in T; \quad (15)$$

$$v_k(t) - \underline{v}_k \geq 0 \perp \underline{\mu}_k^v(t) \geq 0; \forall k \in K, \forall t \in T; \tag{16}$$

$$-v_k(t) + \bar{v}_k \geq 0 \perp \bar{\mu}_k^v(t) \geq 0; \forall k \in K, \forall t \in T; \tag{17}$$

$$P_{sb}(t) - \underline{P}_{sb} \geq 0 \perp \underline{\mu}_{sb}(t) \geq 0; \forall t \in T; \tag{18}$$

$$-P_{sb}(t) + \bar{P}_{sb} \geq 0 \perp \bar{\mu}_{sb}(t) \geq 0; \forall t \in T; \tag{19}$$

$$P_{dg_i}(t) - \underline{P}_{dg_i} \geq 0 \perp \underline{\mu}_i^{dg}(t) \geq 0; \forall t \in T, \forall i \in I; \tag{20}$$

$$-P_{dg_i}(t) + \bar{P}_{dg_i} \geq 0 \perp \bar{\mu}_i^{dg}(t) \geq 0; \forall t \in T, \forall i \in I; \tag{21}$$

$$\lambda_k(t) \text{ free}. \tag{22}$$

The set of equation (9)-(22) represents a mathematical program with equilibrium constraint (MPEC) associated with each DG unit. There are as many MPEC as the number of DG units in the system.

### 2.3 The compact Form

To represent a step-by step of the proposed framework, and for the sake of the simplicity, the formulations are first rewritten in a compact form, where subscripts and time index are discarded.

First, the lower level problem is replaced by its KKT conditions. Note that the lower level problem refers to the Disco optimization problem i.e. (3)-(8). The compact form is presented as follows:

$$\text{Min} \quad f_{DisCo}(\mathbf{x}, \mathbf{w}) \tag{23}$$

subject to

$$Eq(w) = 0 : \lambda \tag{24}$$
$$0 \leq Ineq(w) : \mu \geq 0 \tag{25}$$

where the equality and inequality constraints are denoted as *Eq* and *Ineq*.

Notice that the constraints related to DisCo problem are Eq and Ineq. It should be remembered that the notation in (24) and (25) just show that λ and μ are the dual variables of these constraints.

The vector *x* and *w* include all decision variables of each independent DG unit and the DisCo such that $w = (P_{dg}, P_{sb}, V)$ and $X_i = (\alpha_i)$. The compact form of the KKT conditions is:

$$\nabla_w f_{DisCo}(x, w) - \nabla_w^T Eq(w) * (\lambda) - \nabla_w^T Ineq(w) * (\mu) = 0 \tag{26}$$
$$Eq(\mathbf{w}) = 0 \tag{27}$$
$$0 \leq Ineq(\mathbf{w}) \perp \boldsymbol{\mu} \geq 0 \tag{28}$$
$$\boldsymbol{\lambda} \quad \text{sign-free}. \tag{29}$$

Note that the decision variables of independent DG units are considered as parameters in the Disco problem. Assuming that $y_1 = (w, \lambda)$ and the following functions:

$$h_e(x, \mu, y_1) = \begin{cases} \nabla_w f_{DisCo}(x, y_1) - \nabla_w^T Eq(y_1) * (\lambda) - \nabla_w^T Ineq(y_1) * (\mu) \\ Eq(w) \end{cases} = 0 \tag{30}$$

$$h_{in}(\mathbf{y_1}) = \{Ineq(\mathbf{w})\} \geq 0. \tag{31}$$

By adding slack variables to (31), expressions (26)-(29) can be rewritten as:

$$h_e(\mathbf{x}, \boldsymbol{\mu}, \mathbf{y_1}) = 0 \tag{32}$$

$$h_{in}(\mathbf{y_1}) - \mathbf{s} = 0 \tag{33}$$

$$0 \leq \mathbf{s} \perp \boldsymbol{\mu} \geq 0. \tag{34}$$

Note that expressions (32)-(34) is the appropriate form of the formulations (23)-(25) and so they can be substituted in each DG optimization problem instead of (23)-(25).

For the independent DG $i$, other independent DGs decisions are expressed by $x_{-i}$. Therefore, each DG problem ((2)-(8)) can be written as follows:

$$\begin{pmatrix} Max \quad f_i(x_i, y_1) \\ s.t \\ h_e(x_i, \mu, y_1; x_{-i}) = 0 \\ h_{in}(y_1) - s = 0 \\ 0 \leq s \perp \mu \geq 0 \end{pmatrix} \tag{35}$$

Each DG problem is an MPEC because there are Disco complementarity constraints into their problems. This means that each DG anticipates how the Disco will react to its contract price offer. Considering all the MPEC associated to the DG units, the problem becomes an EPEC which is solved as expressed in the next section.

## 3 THE SOLUTION METHOD

### 3.1 NLP Formulation

The one attractive way to solve EPEC problem is to declare it as an NLP formulation and then use commercially available software like GAMS to solve the problem. The NLP formulation we used is similar to the one presented in [21]. Given the formulation (35) for all DG units and that the complementarity conditions $0 \leq s \perp \mu \geq 0$ can be reformulated in a nonlinear form $-\mu^T s \geq 0$, $\mu, s \geq 0$, each DG problem can be reformulated as a nonlinear form:

$$\begin{pmatrix} Max \quad f_i(x_i, y_1) \\ s.t \\ h_e(x_i, \mu, y_1; x_{-i}) = 0 \quad : \bar{\mu}_i \\ h_{in}(y_1) - s = 0 \quad : \underline{\mu}_i \\ -\mu s \geq 0 \quad : \phi_i \\ s \geq 0 \quad : \sigma_i \\ \mu \geq 0 \quad : \psi_i \end{pmatrix} \tag{36}$$

Then, each MPEC (optimization problem (36)) is replaced by its strong stationarity conditions (which is equivalent to its KKT conditions). Therefore the optimization problem of DG$i$ (optimization problem (36)) can be represented as follow:

$$-\nabla_{x_i} f_i - \nabla_{x_i}^T h_e \bar{\mu}_i = 0; \tag{37}$$

$$-\nabla_{\mu} f_i - \nabla_{\mu}^T h_{in} \underline{\mu}_i - \nabla_{\mu}^T h_e \bar{\mu}_i + s\phi_i - \psi_i = 0; \tag{38}$$

$$-\nabla_{y_1} f_i - \nabla_{y_1}^T h_e \bar{\mu}_i - \nabla_{y_1}^T h_{in} \underline{\mu}_i = 0; \tag{39}$$

$$\underline{\mu}_i + \mu \phi_i - \sigma_i = 0; \tag{40}$$

$$h_e(x, \mu, y_1) = 0; \tag{41}$$

$$h_{in}(y_1) - s = 0; \tag{42}$$

$$-\mu s \geq 0 \quad \perp \quad \phi_i \tag{43}$$

$$s \geq 0 \quad \perp \quad \sigma_i \tag{44}$$

$$\mu \geq 0 \quad \perp \quad \psi_i \tag{45}$$

The complementarity conditions of (36) are expressed as follows:

$$\begin{pmatrix} s \geq 0 \perp \mu \geq 0 \\ \mu \geq \perp \psi_i \geq 0 \\ s \geq 0 \perp \sigma_i \geq 0 \\ \phi_i \geq 0 \end{pmatrix} \tag{46}$$

The NLP formulation problem (36) is derived by replacing the complementarity constraints in (37)-(46) by the equivalent nonlinear form.

As several independent DGs are participated in the Disco supply basket, the complete NLP formulation of the problem is thus obtained by combining all the strong stationarity conditions of each DG units as (47)-(55). Note that the NLP formulation aims to provide a feasible solution for (36) by minimizing the complementarity constraints. For this purpose, first all the complementarity constraints are substituted by $-a^T b \geq 0$, $a, b \geq 0$, then the objective function is defined as a minimization of all the $a^T b$ [21]. This formulation can be written as follows:

$$C_{pen} = Min \sum_i (\sigma_i^T s + \psi_i^T \mu) + \mu^T s \tag{47}$$

Subject to:

$$-\nabla_{x_i} f_i - \nabla_{x_i}^T h_e \bar{\mu}_i = 0, \forall i \in I; \tag{48}$$

$$-\nabla_\mu f_i - \nabla_\mu^T h_{in} \underline{\mu}_i - \nabla_\mu^T h_e \bar{\mu}_i + s\phi_i - \psi_i = 0, \forall i \in I; \tag{49}$$

$$-\nabla_{y_1} f_i - \nabla_{y_1}^T h_e \bar{\mu}_i - \nabla_{y_1}^T h_{in} \underline{\mu}_i = 0, \forall i \in I; \tag{50}$$

$$\underline{\mu}_i + \mu \phi_i - \sigma_i = 0, \forall i \in I; \tag{51}$$

$$h_e(x, \mu, y_1) = 0; \tag{52}$$

$$h_{in}(y_1) - s = 0; \tag{53}$$

$$\psi_i, \sigma_i, \phi_i \geq 0; \tag{54}$$

$$x, \mu, s \geq 0; \tag{55}$$

Note that lower-level problem is nonconvex and hence, the KKT conditions are not sufficient for guaranteeing the optimality of the solution. The NLP formulation (47)–(55) aims to provide a feasible solution by leading complementarity conditions to zero while enforcing all the original system constraints. Since the equilibrium conditions are nonconvex, any solution set with Cpen=0 has to be tested to ensure it corresponds to a local maximum (see also [22, 23]).

To verify the optimality of solution, two different techniques are available. The first technique is to use the diagonalization technique which itself is based on the Gauss-Seidel method. However, the diagonalisation technique does not proof that the solution is optimal.

The second technique, however, definitely guarantees that the lower level decision variables are optimal. In the second technique, the DisCo problem is solved based on EPEC results. In such technique, all contract prices obtained by NLP formulation are put into the DisCo optimization problem. Then DisCo optimization problem is implemented by NLP solvers. In this way all the decision variables of DisCo such as the amount of energy which should be purchased from DGs and wholesale electricity market are obtained. The profits of DGs are also calculated.

If the decision variables of DisCo are the same as the ones obtained by NLP formulation, then the profits of DGs will be the same and the EPEC solution is optimal. Both techniques have been used in this paper to check the optimality of EPEC solution.

### 3.2 Simple example

This section extensively discusses the application of the proposed framework to a simple study. In particular, all DG and DisCo variables and constraints as well as associated dual and complementarity variables have been expressed. NLP formulation and the associated variables and constraints have been demonstrated for the simple study to comprehensively understand the proposed framework and the associated variables and constraints.

A 3-Bus test system, presented in Fig. 2, was used as a simple example in this section to clearly demonstrate how the proposed method works. The data associated with the 3-Bus test system is provided in the Appendix. SE in Fig. 2 refers to the substation through which the energy purchased by DisCo flows into the network. It is assumed that two DG units, DG1 and DG2, was installed on bus 2 and bus 3, respectively.

**Figure 2:** 3-Bus Distribution Test System.

The DisCo problem formulation and associated KKT conditions are represented in (3)-(8) and (10)-(22), receptively. As stated previously, KKT conditions are inserted in each DG problem producing an MPEC.

Now, we aim to present how NLP formulation is constructed by combining the optimization problem of all DG units. First, considering the MPEC formulation (9)-(22), the slack variable *s,* from the set of S, is added to its corresponding inequality constraint. For the sake of simplicity, the time index and the line flow constraints are removed we only consider the problem for one time period. Therefore, the MPEC problem for DG1 is presented as follow:

For DG1, the objective function is:

$$Max \ (\alpha_1 - c_1)P_{dg_1} \tag{56}$$

The DisCo problem formulation is represented in (3)-(8). KKT conditions associated with DisCo problem are represented in (10)-(22) which are inserted in DG1 problem as follows:

$$\nabla_{v_k} \to (-\lambda_k + \bar{\mu}_{kl} - \underline{\mu}_{kl}) \sum_{l \in L} \frac{(2v_k - v_l)}{z_{kl}} + \bar{\mu}_k^v - \underline{\mu}_k^v = 0; \qquad \forall k \in K = \{1,2,3\}; \tag{57}$$

$$\nabla_{P_{dg_i}} \to \alpha_i + \lambda_k + \bar{\mu}_i^{dg} - \underline{\mu}_i^{dg} = 0; \qquad \forall i \in \{1, 2\}; \tag{58}$$

$$\nabla_{P_{sb}} \to \beta + \lambda_k + \bar{\mu}_{sb} - \underline{\mu}_{sb} = 0; \tag{59}$$

It should be remembered that, as shown in equations (5) to (8), $\bar{\mu}_1^v, \bar{\mu}_2^v, \bar{\mu}_3^v, \underline{\mu}_1^v, \underline{\mu}_2^v, \underline{\mu}_3^v$ are the complementary variables associated with the bus voltage limits, $\underline{\mu}_{sb}, \bar{\mu}_{sb}$ are those associated with DisCo power pur-chase limit at the substation and $\bar{\mu}_1^{dg}, \bar{\mu}_2^{dg}, \underline{\mu}_1^{dg}, \underline{\mu}_2^{dg}$ are the complementary variables associated with the DG generation limits.

Bus power flow constraint for the system is presented as:

$$-P_{gk} + P_{dk} + \sum_{l \in L} \frac{v_k(v_k - v_l)}{z_{kl}} = 0; \quad \forall k \in K = \{1,2,3\}; \tag{60}$$

Inequality constraints, equations (5) to (8) are turned into the equality constraints by adding 12-variable slack vector *S:*

$$\bar{v}_k - v_k - s_{k\prime} = 0 \ ; \forall k \in K = \{1,2,3\}; \forall k' \in K' = \{1,2,3\}, K' \subset S; \tag{61}$$

$$v_k - \underline{v}_k - s_{k\prime\prime} = 0 \ ; \forall k \in K; \forall k'' \in K'' = \{4,5,6\}, K'' \subset S; \tag{62}$$

$$\bar{P}_{dg_i} - P_{dgi} - s_{k\prime\prime\prime} = 0 \ ; \forall i \in I = \{1,2\}, \forall k''' \in K''' = \{7,8\}, K''' \subset S; \tag{63}$$

$$P_{dgi} - \underline{P}_{dgi} - s_{k\prime\prime\prime\prime} = 0 \ ; \forall i \in I = \{1,2\}, \forall k'''' \in K'''' = \{9,10\}, K'''' \subset S; \tag{64}$$

$$P_{sb}(t) - \underline{P}_{sb} - s_{k\prime\prime\prime\prime\prime} = 0 \ ; \forall k''''' \in K''''' = \{11\}, K''''' \subset S; \tag{65}$$

$$-P_{sb}(t) + \bar{P}_{sb} - s_{k\prime\prime\prime\prime\prime\prime} = 0 \ ; \forall k'''''' \in K'''''' = \{12\}, K'''''' \subset S; \tag{66}$$

For the sake of simplicity, $\bar{\mu}_1^v, \bar{\mu}_2^v, \bar{\mu}_3^v, \underline{\mu}_1^v, \underline{\mu}_2^v, \underline{\mu}_3^v, \underline{\mu}_{sb}, \bar{\mu}_{sb}, \bar{\mu}_1^{dg}, \bar{\mu}_2^{dg}, \underline{\mu}_1^{dg}, \underline{\mu}_2^{dg}$ was substituted by $\mu_1, \mu_2, \mu_3, \mu_4, \mu_5, \mu_6, \mu_7, \mu_8, \mu_9, \mu_{10}, \mu_{11}, \mu_{12}$, respectively. Therefore we have the following constraints:

$$-\mu s \geq 0 \quad ; \{\mu_1 \dots \mu_{12}\}, S = \{s_1 \dots s_{12}\} \tag{67}$$

$$-s \geq 0 \tag{68}$$

$$\mu \geq 0 \tag{69}$$

The objective function (56) subjected to the set of equations (57)-(69) forms the MPEC associated with DG1. A set of equations similar to (56)-(69) can also be derived to form MPEC problem associated with DG2.

As stated in the previous section, DG1 and DG2 problems are turned into the strong stationarity conditions. Combining the strong stationarity conditions, NLP formulation is developed as follows:

The objective function for NLP formulation is presented in (70). The aim of objective function is to push the sum of the product of all complementarity variables into zero. Variables $\psi_i, \sigma_i, \phi_i$ are the vectors for the associated constraints (67)-(69) for each DG unit.

$$C_{pen} = Min \sum_i (\sigma_i^T s + \psi_i^T \mu) + \mu^T s \tag{70}$$

Equation (71) represents derivatives of DG decision variable, i.e. contract price ($\alpha_i$).

$$-\nabla_{\alpha_i} f_i - \nabla_{\alpha_i}^T h_e \bar{\mu}_i = 0, \forall i \in I = \{1, 2\}; \tag{71}$$

In (71), $f_i$ represents $DG_i$ objective function and $h_e$ represents all the equality constraints in MPEC associated with $DG_i$ i.e. (57)-(60). Variable $\bar{\mu}_i$ in (71) is the complementarity variable associated with $h_e$.

Derivative of $\mu$ are presented in (72) in which $h_{in}$ represents all the inequality constraints in MPEC associated with $DG_i$ as presented in (61)-(69). ) $\underline{\mu}_i$ is the complementarity variable corresponding to $h_{in}$.

$$-\nabla_\mu f_i - \nabla_\mu^T h_{in} \underline{\mu}_i - \nabla_\mu^T h_e \bar{\mu}_i + s\phi_i - \psi_i = 0, \forall i \in I; \tag{72}$$

Derivatives of MPEC associated with DG units to $y_1$ is presented in (73) in which $y_1$ is a vector consists of $P_{sb}$, $P_{dg_1}$, $P_{dg_2}$ and also $v_k$ and $\lambda_k$ vectors.

$$-\nabla_{y_1} f_i - \nabla_{y_1}^T h_e \bar{\mu}_i - \nabla_{y_1}^T h_{in} \underline{\mu}_i = 0, \forall i \in I; \tag{73}$$

Set of equation (74) represents the derivative of S={s1, …, s12} for each DG problem.

$$\underline{\mu}_i + \mu \phi_i - \sigma_i = 0, \forall i \in I; \tag{74}$$

Constraints on $h_e$ and $h_{in}$ and the problem variables have been presented in (75)-(78) in which $x$ is the vector of decisions variables for DG units, i.e. $(\alpha_1, \alpha_2)$.

$$h_e(x, \mu, y_1) = 0; \tag{75}$$

$$h_{in}(y_1) - s = 0; \tag{76}$$

$$\psi_i, \sigma_i, \phi_i \geq 0; \tag{77}$$

$$x, \mu, s \geq 0; \tag{78}$$

The formulations (70)-(78) refers to the proposed NLP formulation for EPEC problem which can be solved by NLP solver. If the objective functions become zero then the feasible solution is attained.

The proposed framework was applied to the 3-Bus system. For the sake of simplicity, it was assumed that DisCo can purchase the electricity from a retailer at the fixed price of 60 €/MWh. Note that the retailer price is equal to the DG production cost. Moreover, the system demand is 6 MW which is equally distributed in the system buses and the demand is constant all over the year. Without DG application, system loss is 0.057 MW and DisCo should purchase the system demand as well as system loss, i.e. 6.057 MW, from the retailer.

Two 1MW DG units, DG1 and DG2, have been located in Bus 2 and Bus 3, respectively. The proposed framework has been applied to determine contract price and power of DG units for a yearly electricity purchase contract between DisCo and DG units. Optimal contract price for DG1 and DG2 has been obtained as 60.68 and 61.01 €/MWh, respectively and the contracted power for each DG unit is 1 MW, meaning that DG units are fully contracted. So, competition among DG units has led to different contract prices for DG units.

Table 1 shows the DisCo payment and energy loss with and without DG application.

**Table 1:** Competition Results for Simple Example

It can be seen in Table 1 that, even for a simple case, the competition among DG units has led to decrease in total DisCo payment.

The computation details are presented in Table 2.

**Table 2:** Computation Details for Simple Example

## 4   NUMERICAL RESULTS

The proposed model has been applied to the modified IEEE 34-Bus Distribution Test System [14], presented in Fig. 3. The additional information used in this paper for the IEEE 34-Bus Distribution Test System is provided in Table 10 in Appendix.

**Figure 3:**   IEEE 34-Bus Distribution Test System [15].

Total system load has been scaled such that the maximum system load is 10 MW. The annual load duration curve of the system is assumed to be composed of five equal periods each cover 20 percent of the year as the one shown in Fig. 4.

For each of the five load periods, the wholesale energy market prices in €/MWh at the substation (bus 1) are given in Fig. 4. It can be seen in Fig. 4 that market price in 40 percents of the time is not less than 70 €/MWh.

**Figure 4:** Annual System Load Duration Curve and Market Price.

Four dispatchable DG units each with a capacity of 1 MW was assumed denoted as DG1, DG2, DG3 and DG4. The production cost of all DG units was assumed similar and equals to 60 €/MWh. The reason for assuming similar production cost for all DG units is that the current paper is aimed to evaluate the effect of competition and location on DG contract pricing problem. If DG units have identical production cost, the impact of competition on the contract price of DG units can be easily identified through analyzing the difference in the profit of DG units.

The proposed framework has been applied to the Modified IEEE 34-Bus test system to determine a fixed one-year optimal DG contract price. For the sake of simplicity, it is assumed that the load at each period is equally divided among the nodes. All simulations were carried out on a laptop computer with a 1.1 GHz processor and 4 GB of RAM memory. The proposed framework was implemented in GAMS environment and solved using IPOPT solver.

Two scenarios have been studied:
- Scenario A: only two DG units are considered.
- Scenario B: The number of DG units increased to four.

Several cases are simulated as presented next.

## 4.1 Scenario A

In this scenario, DG1 was located in bus 17 and DG2 was applied in bus 24. Two cases have been studied for Scenario A:
- Case 1: This case represents the base case.
- Case 2: This case maintains the conditions in Case 1 and further assumes the production cost of DG1 increased to 70 €/MWh.

The proposed framework has been applied and the results were given in Table 3 where the optimal contract price, annual energy produced by DG and their profits are presented. Note that the contract price is fixed during the year.

**Table 3:** Results of Scenario A.

Regarding the contracted DG power, it was observed that, for cases 1, both DG1 and DG2 sell the rated power (1 MW) to DisCo during 40 percent of the time when the market price is higher than 62 €/MWh and the contracted power is zero in other times. This shows that DisCo take advantage of DG generation in high market price periods.

It can be seen in Table 3 that for Case 1, contract price of DG units is well above the DG generation cost. In particular, DG2 has higher profit in both cases and the DisCo purchases the energy from DG2 at a higher price. This is because DG2 is located at a farther bus w.r.t the substation and can reduce energy loss more effectively.

By increasing the production cost of DG1 in Case 2, it is observed that DG1 sells the energy at a higher price than in Case 1. However, as DG1 increased the contract price, DisCo reacted to this action and decreased the amount of energy purchased from DG1 to 1752MWh. Therefore, the profit of DG1 decreased. It is interesting to note that DG2 also changes the contract price to the higher level in Case 2 compared to Case 1. As it is seen in Table 2, the profit of DG2 increases by 13.34% in Case 2 compared to Case 1 as a response to increase in the production cost of DG1 in Case 2. This shows that the proposed framework can consider the benefits of each player properly so that the interaction between competing DG units can be correctly studied.

Table 4 shows the DisCo payment and annual energy losses with and without DG units.

**Table 4:** Outcomes in Scenario A.

It can be seen in Table 4 that, by dispatching DGs, DisCo cost and annual energy loss were reduced by 1.83% and 31.39%, respectively, in Case 1 and 0.98% and 26.14% in Case 2 compared to the case where no DG unit was used. It can also be seen in Table 3 that DisCo energy purchase at higher market prices was also reduced when DG used.

So, DG application is beneficial to DisCo and resulted to reduction in DisCo payment. On the other hand, DisCo purchases the energy from DG units at a price higher that DG production cost. It can be interpreted that DisCo shared a part of benefit from DG application with DG units by setting DG contract price higher than DG production cost. Competition among DG units has led to higher price for DG2 than DG1, as DG2 is more beneficial to DisCo than DG1.

### 4.2 Scenario B

To show the efficacy of the proposed framework, more than two DG units were considered in Scenario B. In this scenario, two cases were studied as described below.
- Case 3: Three DG units were considered: DG1 at bus 17, DG2 at bus 24, and DG3 at bus 11.
- Case 4: It maintains the condition in Case 3 and assumes that DG4 is located in bus 33.

Case3 and Case 4 have been analyzed using the proposed framework. Table 5 presents the results. Contracted DG generation was the same as those obtained for Case1 and Case 2 in Scenario A.

**Table 5:** DG Results for Different Cases in Scenario B.

It can be observed that profit of DG units has been decreased with penetration of more DG units. Compared to Case 1, the profit of DG1 and DG2 decrease by 2.48% and 2.51% in Case 3, and by 5.42% and 6.16% in Case 4. It also can be seen that DG2 has the highest contract price even when DG4 was installed on a farther bus. This is due to the fact that DG2 plays more important role in the power flow in the distribution network and it can receive higher contract price. So, it can be concluded that the DG contract price depends on DG usefulness for the DisCo rather than the DG distance from the substation.

Comparing the results presented for Case 4 and Case 1, it is interesting to note that the profit of DG2 reduced more than that of DG1 with the increase in the number of DG units. In this study, DG2 gains the highest profit while DG3 attains the lowest profit. DisCo Payments and losses are shown in Table 6.

**Table 6:** Disco Results for Different Cases in Scenario B.

As seen from Table 6, the DisCo payments decrease by 0.92% and 2.33% in Case 3 and Case 4 compared to Case 1. Note that as the DG penetration level increases, the DisCo cost decreases accordingly.

### 4.3 Impact of Competition

In this section the impact of modelling the competition is investigated and its effect on the price and costs are examined.

A case was simulated referred to as OW1 which hold the same conditions given for Case 1. However, we assumed that DG units belong to a single owner and do not compete with each other. The DG owner thus will try to maximize the profit defined below:

$$\underset{\alpha_i}{Max} \sum_{i \in I} \sum_{t \in T} (\alpha_i - c_i) p_{dg_i}(t) \tag{79}$$

From formulation (79), it is seen that the DG owner tries to maximize the total profit equals to the summation of all DGs' profits. However, in the proposed framework (see formulation (2)) each DG unit tries to maximize its own profit considering the action of other DGs and reaction of DisCo. It should be noticed that with the above objective function and assumptions, an MPEC formulation is obtained to calculate DG contract price as the model proposed in [14]- [16]. The resulting MPEC has been solved using the SNOPT solver in GAMS environment and the results have been obtained. Table 7 compares the results obtained for the case OW1 with that obtained in Part 4.1 for Case1.

**Table 7:** Competition Impact Evaluation Results

According to Table 7, compared to Case 1, the contract price and profit of DG2 increased in OW1 while the profit decreased for DG1. Results in Table 6 show that the difference between the profits of DG2 and DG1 is 4,650 € in Case 1, while it increased to 22,128 € in OW1. In OW1, the DG owner's objective is to attain the best profit and it would raise the price as it is the sole choice for the DisCo. As expected, the lower DisCo payments and energy loss take place in Case 1 instead of OW1. In fact, the competition among DG units in Case 1 has led to more benefits for the DisCo and it also causes to decrease the energy loss by 7.69%. Note that although the total profit of DG units in OW1 is higher than that in Case1, some DG units can gain more profit by deviating from their contract prices in OW1. For example, DG1 can increase its profit by 12.73% if it changes its contract price to the price calculated in Case1.

Results obtained in this section showed that the competition among DG units permits the DisCo to minimize the cost. It also resulted in smaller energy loss and help increase energy efficiency. For instance, in Case 1, DG1 and DG2 are dispatched properly so that the DisCo payments and energy losses decreased and no player gains more profits by deviating from his contract price. Moreover, DG2 as a player who has more positive impact on distribution network and energy losses gained more profits. Therefore, modelling the competition can capture the benefits of DG to the distribution network properly and bring more profit for DG2. In contrast, neglecting the competition may result in an unreasonable solution for the contract pricing problem. For example, as observed in OW1, DG1 has to reduce his profits in favour of DG2 for the sake of global optimality (formulation (79)).

On the other hand, results presented in this section showed that the competition among DG unit must be correctly modelled to determine competitive contract price. Otherwise, the prices that are calculated by ignoring the competition will have a considerable drift with the correct competitive prices.

## 5 OPTIMALITY EVALUATION

As previously indicated, due to the nonconvexity of the lower level problem (3)-(8), the optimality of the solution should be verified. In this section, the problem is solved using two other techniques and the results are compared with those presented in Section 4 which are calculated using NLP technique. The techniques used to verify the results are called the diagonalisation technique and post-optimality test technique.

Diagonalization technique is a technique to solve EPEC that has been widely used [23] and [24]. This technique was used and different starting points were tested to verify the optimality but only one Nash point was found. It should be noted that the diagonalization techniques were implemented in GAMS environment and solved using the IPOPT solver. In all cases, the solutions found by the NLP approach were very close to the solutions of the diagonalization method and differences were negligible. For instance, the difference between the DG contract prices using these methods are around $10^{-4}$. The computational details of the proposed framework and diagonalization technique are presented in Table 8.

**Table 8:** Computation Details

The accuracy in Table 8 is defined for NLP as the values of objective function (47). For the diagonalization technique, accuracy is defined as the difference between DG contract prices in two consecutive iterations. Notice the large problem size and the longer computing time for the NLP approach. In the case of the diagonalization technique, two and three sweeps are needed to find the solution for Case 3 and Case 4, respectively. Note that although the problem size and computing time were reduced in the diagonalization technique, cycling issue is the major drawback of this technique [25].

In the diagonalisation technique, each DG problem will be solved separately, and by iteration and updating the decision variables of DG units the optimal solution will be obtained. That is why the number of variables, constraints, iteration and CPU second per problem in Table 8 is different from those associated with the complete problem.

However, NLP technique considers all KKT conditions of DG problems simultaneously and only one optimization problem will be solved. It means the number of variables/constraints per problem in Table 8 for the NLP formulation is equal to those associated with whole problem. This would be the same for CPU seconds/iteration per problem and total amount.

Once the results have been verified using the diagonalization technique, the post-optimality test technique was also used. This technique is based on the Nash equilibrium definition which states that no player will be better paid off if it changes the strategy unilaterally.

Remembering that DG units act as the players of the Nash game, the post-optimality test works as follows: First, all decision variables of the DG units are fixed at their optimal values obtained using the proposed framework and are substituted in the DisCo problem (3)-(8). Second, the contract price of one DG unit is changed around its neighbourhood and the DisCo problem is solved and the profit of DG units is calculated. If the profit of the DG unit with changed contract price is smaller than the one obtained by the proposed framework, it means that the contract price calculated by the proposed framework is optimal. This test is carried out for all DG units. Finally, if the optimality of the contract price of all DG units is assured, it indicates that the solution corresponds to an optimal point and is thus the Nash equilibrium.

Using the second technique, the profits of all DG units were computed by changing the contract prices around those obtained from the NLP technique. Fig. 5 plots the profits of DG1 in Case1 for different contract prices.

**Figure 5:** DG1 Profit for Different Contract Prices Near The Optimal Value.

It can be seen in Fig. 5 that the highest profit of DG1 occurs at the point which obtained by the proposed framework. Similarly the contract prices of DG units for different scenarios are checked. Verification tests showed that all DGs obtain no higher profits by deviating from the solutions of the NLP technique, meaning that the solutions obtained from the NLP technique are indeed optimal.

So, the optimality of NLP solution has been successfully checked by two different techniques.

## 6   PRACTICAL IMPLICATIONS

The performance of the proposed framework in determining DG contract price considering competition has been carefully demonstrated and evaluated in the previous sections. The proposed framework can be easily used by DisCo or DG owners to propose a contract price with sound and logical economic and technical basis.

While the proposed framework can be easily and extensively used, there are some issue regarding practical implication of the proposed framework. These issue are addressed in this section.

### 6.1. Introduction of Retailer Company

The proposed framework is based on the assumption that DG sells the electricity merely to DisCo. However, in some market structures, retailer companies can buy the DG generation. A question is, then, how would be the applicability of the proposed framework in the presence of retailer companies.

The proposed framework, however, can be easily extended to consider the competition between DisCo and retailers companies to buy DG generation. It is only needed to add the optimization problem of the retailer companies in the lower level problem. In such case, there would be several lower level problems which act as constraints in each DG problem. Substituting the KKT conditions associated with the problems, a NLP formulation for the problem shall be achieved that can be easily solved to determine the DG contract price with DisCo and retailer company.

However, advanced market structures with retailer also bring other options for DG owners such as participating in the demand response programs or peak shaving services. The authors are working on a research project to obtained suitable methods to determine multiple-contract pricing for DG units including retailer contract in an advanced market structure.

### 6.2. Use of Prices

It is evident that the current practice for DG pricing does not necessarily follow the proposed pricing algorithm. Electricity generation of DG is, generally, purchased by the electric utility at the negotiated prices. In some markets, such as Iranian power market, long term fixed-price power purchase contract is established between the DG and the utility. DisCo may choose any basis to determine the contract price and it can even cross subsidize between its various activities. So, there is no guarantee that the contract prices obtained by the proposed framework will be applied.

However, an important question is how to set the electricity purchase price which is technically and economically sound and justifiable. The question becomes more important if different independent DGs are located in the network, each of which has different advantage to the electric utility. A technically sound framework will be needed in near future to determine optimal contract price. The paper proposes a framework to determine the optimal DG contract price and can be used by either DG owners or DisCo to determine a techno-ecumenically justifiable contract price.

## 7   CONCLUSION

This paper has proposed a framework for optimal contract pricing of dispatchable DG units in the distribution system considering the competition between DG units. A multi-DG equilibrium problem was developed to represent the interaction of DG units. Using a multi-period bilevel optimization framework, each DG problem was formulated as an MPEC problem which was transformed into an EPEC model considering all DG units. By mathematical reformulation, the resulting EPEC problem was represented as a nonlinear model and solved using NLP solvers. In addition, two heuristic tests were proposed and applied to verify the optimality of the solution.

The proposed framework was applied to the modified IEEE 34-Bus Test System and the impact of DG location, load demand and the capacity of DGs on the contract pricing problem were studied. It was shown that by considering the competition between the DG units, the DisCo will gain more benefits by decreasing the operating cost of the system. Moreover, it was observed that considering the competition gives higher profit to more beneficial units and can adjust the payments according to the importance of each unit to the distribution network. On the other hand, neglecting the competition between the DG units may results in an unrealistic solution in which the profit of some DG units may be sacrificed in favour of other DGs' profits.

The proposed framework captures the competition between DG units properly and can be easily extended to contain any number of DGs. However, the proposed approach should be extended to include non-dispatchable DG units as well as the contract between DG units and retailers.

# 8   APPENDIX

Table 9 represents the data associated with 3-Bus Test system. Note that, in Table 9, only the necessary data for solving the problem is provided. Thus the impedance magnitude is mentioned instead of resistance and reactance of the lines. The system load was also divided equally among the nodes.

**Table 9:** 3-Bus Distribution Test System Data (Based on 10 MVA base)

The extra data used for IEEE 32-Bus distribution Test system are provided in Table 10.

**Table 10:** Extra Data for IEEE 34-Bus Distribution Test System (Based on 10MVA base)

**Ashkan Sadeghi Mobarakeh** received the B.S. degree in electrical engineering from Amirkabir University of Technology (Tehran Polytechnic), Tehran., Iran, in 2010. He received M.Sc. in Energy Systems Engineering at Sharif University of Technology in 2012. Recently, He has been accepted by the University of California, Riverside to start his PhD. His main fields of research are game theory application in power market and distributed generation.

**Abbas Rajabi-Ghahnavieh (M'08)** was born in Iran in 1981. He received his B.Sc. in electrical engineering from Isfahan University of Technology, Isfahan, Iran in 1999. He received M.Sc. in electrical engineering from Sharif University of Technology(SUT), Tehran, Iran in 2003. Dr. Rajabi-Ghahnavieh has obtained his Ph.D. in electrical engineering in joint Ph.D. program between SUT and Grenoble INP, Grenoble, France in 2010. He joint Department of Energy engineering at SUT in 2010 as assistant pro-fessor. He has been a visiting fellow to the University of New South Wales, Sydney, Australia on 2012. His research area includes reliability evaluation of engineering systems and smart grid studies.

**Hossein Haghighat (M'08)** received the Ph.D. degree in electrical engineering from Tarbiat Modares University, Tehran, Iran, in 2007. His research interests include electricity market operation and optimization.

**Figure 1:** Schematic of the Proposed Model.

**Figure 2:** 3-Bus Distribution Test System.

**Figure 3:** IEEE 34-Bus Distribution Test System [15].

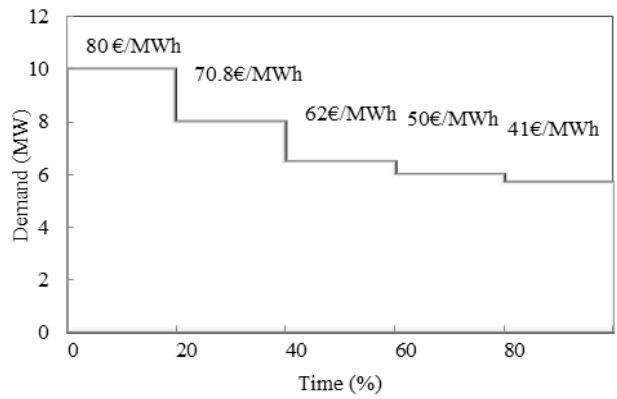

**Figure 4:** Annual System Load Duration Curve and Energy Market Price.

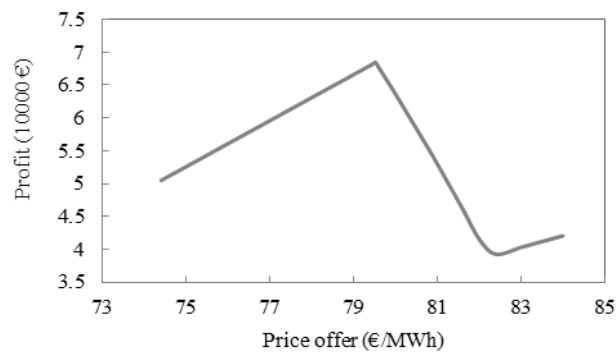

**Figure 5:** Profit of DG1 for Different Contract Prices Near The Optimal Value.

**Table 1:** Competition Results for Simple Example

| Index | With DG | Without DG |
|---|---|---|
| Energy Loss (MWh) | 0.01 | 0.057 |
| Total DisCo Payment (€) | 3,173,660.4 | 3,183,559.2 |
| DisCo payment to Retailer (€) | 2107656 | 3,183,559.2 |
| DisCo Payment to DG1 (€) | 531556.8 | - |
| DisCo Payment to DG2 (€) | 534447.6 | - |

**Table 2:** Computation Details for Simple Example

| Index | NLP |
|---|---|
| Number of Variables of the DisCo Problem | 12 |
| Number of Constraints of DisCo Problem | 15 |
| Number of Variables of Each DG Problem | 28 |
| Number of Constraints of Each DG Problem | 33 |
| Number of Variables NLP Formulation | 148 |
| Number of Constraints NLP Formulation | 77 |
| Total CPU Seconds | .53 |
| Total Solver Iterations | 69 |
| Accuracy | 1.6E-9 |

**Table 3:** Results of Scenario A.

| Index | DG | Case 1 | Case 2 |
|---|---|---|---|
| Contract Price (€/MWh) | 1 | 79.53 | 94.65 |
| | 2 | 80.85 | 83.64 |
| Annual Generated Energy (MWh) | 1 | 3504 | 1752 |
| | 2 | 3504 | 3504 |
| Profit (€) | 1 | 68,433 | 43,186 |
| | 2 | 73,083 | 82,834 |

**Table 4:** Outcomes in Scenario A.

| Index | | With DG | | Without DG |
|---|---|---|---|---|
| | | Case 1 | Case 2 | Case 1&3 |
| Energy Loss (MWh) | | 3,239 | 3,487 | 4,721 |
| Total DisCo Payment (€) | | 4,269,480 | 4,308,010 | 4,348,877 |
| DisCo Payment to DG1 (€) | | 278,539 | 165,827 | - |
| DisCo Payment to DG2 (€) | | 283,298 | 293,074 | - |
| DisCo Payment to Market (€) | | 3,707,643 | 3,849,107 | 4,348,877 |
| DisCo Energy Purchased from Market at different price levels (MWh) | 80 €/MWh | 1,196,165 | 1,196,165 | 1,545,991 |
| | 70.8 €/MWh | 777,321 | 918,786 | 1,068,729 |
| | 62 €/MWh | 748,666 | 748,666 | 748,666 |
| | 50 €/MWh | 553,462 | 553,462 | 553,462 |
| | 41 €/MWh | 432,028 | 432,028 | 432,028 |

**Table 5:** DG Results for Different Cases in Scenario B.

| Index | DG | Case 3 | Case 4 |
|---|---|---|---|
| Contract Price (€/MWh) | 1 | 77.56 | 75.22 |
| | 2 | 78.82 | 75.87 |
| | 3 | 76.44 | 74.58 |
| | 4 | - | 75.82 |
| Annual Generated Energy (MWh) | 1 | 3504 | 3504 |
| | 2 | 3504 | 3504 |
| | 3 | 3504 | 3504 |
| | 4 | - | 3504 |
| Profits (€) | 1 | 61,530 | 53,331 |
| | 2 | 65,924 | 55,608 |
| | 3 | 57,605 | 51,088 |
| | 4 | - | 55,433 |

**Table 6:** DisCo results for different Cases in scenario B.

| Index | | Case 3 | Case 4 |
|---|---|---|---|
| Energy Loss (MWh) | | 2,829 | 2,387 |
| Total DisCo Payment (€) | | 4,227,813 | 4,170,388 |
| DisCo Payment to DG1 (€) | | 271,770 | 263,291 |
| DisCo Payment to DG2 (€) | | 276,185 | 265,849 |
| DisCo Payment to DG3 (€) | | 267,846 | 261,328 |
| DisCo Payment to DG4 (€) | | - | 265,673 |
| DisCo Payment to Market (€) | | 3,412,012 | 3,114,247 |
| Energy Purchased from Market at | 80 €/MWh | 1,036,247 | 873,984 |
| | 70.8 €/MWh | 641,609 | 506,107 |

| different price lev- | 62€/MWh | 748,666 | 748,666 |
| els (MWh) | 50 €/MWh | 5,53,461 | 5,53,461 |
| | 41 €/MWh | 432,028 | 432,028 |

**Table 7:** Competition Impact Evaluation Results

| Index | DG | Case 1 | OW1 |
|---|---|---|---|
| Contract Price (€/MWh) | 1 | 79.53 | 94.65 |
| | 2 | 80.85 | 83.64 |
| Profits (€) | 1 | 68,433 | 60,707 |
| | 2 | 73,083 | 82,834 |
| Total profits (€) | - | 141,925 | 143,541 |
| DisCo Payments (€) | - | 4,269, 480 | 4,303,220 |
| Losses (MWh) | - | 3,239 | 3,488 |

**Table 8:** Computation Details

| Index | Case1 | | Case 3 | | Case 4 | |
|---|---|---|---|---|---|---|
| | NLP | Diag. | NLP | Diag. | NLP | Diag. |
| Number of variables per problem | 4,768 | 1,097 | 6,834 | 1,132 | 8,940 | 1,157 |
| Number of constraints per problem | 2,928 | 1,101 | 4,149 | 1,136 | 5,395 | 1,161 |
| Number of variables total | 4,768 | 4,388 | 6,834 | 6,792 | 8,940 | 13,884 |
| Number of constraints total | 2,928 | 4,404 | 4,149 | 6,816 | 5,395 | 13,932 |
| Total CPU seconds per problem | 4.992 | 5.241 | 13.525 | 6.833 | 78.718 | 17.817 |
| Total solver iterations per problem | 61 | 196 | 113 | 264 | 411 | 633 |
| Total CPU seconds | 4.992 | 20.964 | 13.525 | 27.33 | 78.718 | 106.9 |
| Total solver iterations | 61 | 784 | 113 | 1056 | 411 | 3798 |
| Accuracy | 1.40E-9 | 7.9E-6 | 1.18E-6 | 9.4E-6 | 2.75E-9 | 6.6E-6 |

**Table 9:** 3-Bus Distribution Test System Data (Based on 10 MVA base)

| Item | Value | Unit |
|---|---|---|
| Line 1-2 Impedance Magnitude | 1.236 | p.u. |
| Line 2-3 Impedance Magnitude | 1.144 | p.u. |
| Total System Load | 0.6 | p.u. |
| Active Power Demand at Each Node | 0.2 | p.u. |
| Line Load Limit | 1 | p.u. |
| Max. Substation Active Power | 4 | p.u. |
| Min. Substation Active Power | 0 | p.u. |
| Max. Voltage Magnitude Limit | 1.05 | p.u. |
| Min. Voltage Magnitude Limit | 0.9 | p.u. |
| DG Unit Max. Active Power Limit | 1 | MW |
| DG Unit Min. Active Power Limit | 0 | MW |
| DG Unit Production Cost | 60 | €/MWh |

**Table 10:** Extra Data for IEEE 34-Bus Distribution Test System (Based on 10MVA base)

| Item | Value | Unit |
|---|---|---|
| Total System Load | 1 | p.u. |
| Active Power Demand at Each Node | 0.03125 | p.u. |
| Line Load Limit | 10 | p.u. |
| Max. Substation Active Power | 10 | p.u. |
| Min. Substation Active Power | 0 | p.u. |
| Max. Voltage Magnitude Limit | 1.05 | p.u. |
| Min. Voltage Magnitude Limit | 0.9 | p.u. |
| DG Unit Max. Active Power Limit | 1 | MW |
| DG Unit Min. Active Power Limit | 0 | MW |
| DG Unit Production Cost of DG | 60 | €/MWh |